\documentclass[aps,prl,twocolumn,groupedaddress,floatfix]{revtex4}

\usepackage{graphicx}% Include figure files
\usepackage{dcolumn} % Align table columns on decimal point
\usepackage{bm}      % bold math

\begin{document}

\title{Using photoemission spectroscopy to probe a strongly interacting Fermi gas}
\author{J. T. Stewart}
\author{J. P. Gaebler}
\author{D. S. Jin}
\email[Electronic address: ]{jin@jilau1.colorado.edu} \homepage[URL:
]{http://jilawww.colorado.edu/~jin/}

\affiliation{JILA, Quantum Physics Division, National Institute of
Standards and Technology and Department of Physics, University of
Colorado, Boulder, CO 80309-0440, USA}

%\date{\today} \begin{abstract}
%
%\end{abstract}

 \pacs{??)}

\maketitle

%%Beginning of Text
\textbf{Ultracold atom gases provide model systems in which
many-body quantum physics phenomena can be studied. Recent
experiments on Fermi gases have realized a phase transition to a
Fermi superfluid state with strong interparticle interactions
\cite{Regal2003a, Regal2004a, Chin2004a, Partridge2005,
Zwierlein2005b, Luo2007, Tarruell2008}. This system is a realization
of the BCS-BEC crossover connecting the physics of BCS
superconductivity and that of Bose-Einstein condensation (BEC)
\cite{Eagles1969, Leggett1980, Nozieres1985}. While many aspects of
this system have been investigated, it has not yet been possible to
measure the single-particle excitation spectrum, which is a
fundamental property directly predicted by many-body theories. Here
we show that the single-particle spectral function of the strongly
interacting Fermi gas at $T \approx T_c$ is dramatically altered in
a way that is consistent with a large pairing gap. We use
photoemission spectroscopy to directly probe the elementary
excitations and energy dispersion in the Fermi gas of atoms. In
these photoemission experiments, an rf photon ejects an atom from
our strongly interacting system via a spin-flip transition to a
weakly interacting state. We measure the occupied single-particle
density of states for an ultracold Fermi gas of $^{40}$K atoms at
the cusp of the BCS-BEC crossover and on the BEC side of the
crossover, and compare these results to that for a nearly ideal
Fermi gas. Our results probe the many-body physics in a way that
could be compared to data for high-Tc superconductors
\cite{Perali2002}. This new measurement technique for ultracold atom
gases, like photoemission spectroscopy for electronic materials,
directly probes low energy excitations and thus can reveal
excitation gaps and/or pseudogaps. Furthermore, this technique can
provide an analog to angle-resolved photoemission spectroscopy
(ARPES) for probing anisotropic systems, such as atoms in optical
lattice potentials.}

As interacting quantum systems with highly tunable parameters and
well understood two-body interactions, ultracold atom gases provide
model systems in which to test condensed matter theories. A
challenge for experimenters is to find ways to probe these atom
gases that relate directly to condensed matter ideas and enable
sensitive searches for new phenomena that can advance our
understanding of strongly correlated systems. At a very basic level,
the effect of interactions is a modification of the single-particle
states. As interactions are increased, the single-particle
eigenstates of the non-interacting case become quasi-particles and
phase transitions manifest themselves as qualitative changes to the
excitation spectrum \cite{Fetter2003}, such as the appearance of
energy gaps. The single-particle excitation spectrum can be
predicted by many-body theory and is a fundamental property of any
interacting system.

\begin{figure}
\includegraphics[width=\linewidth]{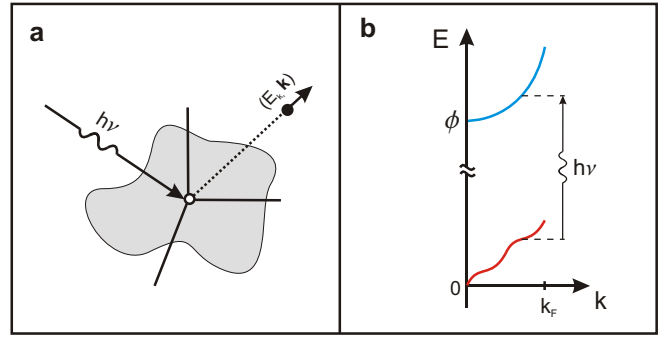}
\caption{\label{arpes_compare} \textbf{Photoemission spectroscopy
for ultracold atom gases}. \textbf{a} In electron PES, one measures
the energy of electrons emitted from solids, liquids, or gases by
the photoelectric effect.  Using energy conservation, the original
energy of the electrons in the substance can be determined.
Similarly, in photoemission spectroscopy for atoms, an rf photon
with energy, $h\nu$, transfers atoms into a weakly interacting spin
state. \textbf{b} The rf photon drives a vertical transition where
the momentum $\hbar k$ is essentially unchanged. By measuring the
energy and momentum of the out-coupled atoms (upper curve) we can
determine the quasiparticle excitations and their dispersion
relation (lower curve). Here $\phi$ is the Zeeman energy difference
between the two different spin states of the atom.}
\end{figure}

For electronic systems, photoemission spectroscopy (PES) provides a
powerful technique to probe the occupied single-particle states
\cite{Damascelli2004}. In a typical PES experiment, electrons are
ejected from a substance through the photoelectric effect, see Fig.
\ref{arpes_compare}a. The photoelectrons are collected, energy and
momentum resolved, and counted to give a spectrum of intensity as a
function of the measured kinetic energy, $\epsilon_k =\hbar^2
k^2/2m$. Here, $\hbar=h/2\pi$, where $h$ is Planck's constant, and
$m$ is the particle mass. By conservation of energy, one can
determine the energy of the original single-particle state, $E_s$,
using
\begin{equation}\label{eq:ConservE}
E_s = \epsilon_k +\phi - h\nu.
\end{equation}
Here, $h\nu$ is the photon energy, $\phi$ is the work function of
the surface, and $E_F-E_s$ is often referred to as the binding
energy \cite{Damascelli2004}.

For ultracold atom gases, radio-frequency (rf) spectroscopy has been
used to probe a strongly interacting Fermi gas \cite{Regal2003a,
Chin2004a, Regal2003b, Gupta2003, Shin2007, Schunck2007,
Schunck2008}. In a typical experiment, a pulse of rf drives atoms
into an unoccupied Zeeman spin state, where they are counted to
yield a spectrum of counts versus rf frequency. To date, the rf
out-coupled atoms have not been energy or momentum resolved.
However, analogous to electron PES, the momentum of the rf photon is
negligible compared to the typical momentum of the atoms and
therefore the momenta of the out-coupled atoms are characteristic of
the original atom states. Eqn. \ref{eq:ConservE} applies to
photoemission spectroscopy of atom gases, by means of
momentum-resolved rf spectroscopy, if one simply replaces the work
function $\phi$ with the Zeeman energy splitting, see Fig.
\ref{arpes_compare}b. The extension of photoemission spectroscopy
from condensed matter to cold Fermi gases was discussed by Dao {\it
et al.} \cite{Dao2007}.

In this paper, we use photoemission spectroscopy, by means of
momentum-resolved rf spectroscopy, to probe an ultracold gas of
fermionic $^{40}$K atoms.  Similar to PES in solids, this
measurement probes the single-particle spectral function, which is
directly related to the single-particle Green's function predicted
by many-body theories \cite{Damascelli2004}. We use this new
technique to probe the Fermi gas near a magnetic-field Fano-Feshbach
resonance where one can tune strong atom-atom interactions to
realize a Fermi superfluid in the region of the BCS-BEC crossover
\cite{Regal2003a, Regal2004a, Chin2004a, Partridge2005,
Zwierlein2005b, Luo2007, Tarruell2008}.

Our Fermi gas consists of $3\times10^5$ $^{40}$K atoms in a mixture
of two spin-states. The gas is confined in an optical dipole trap
and evaporatively cooled to $T/T_F=0.18$, where $T$ is the
temperature, $T_F$ is the Fermi temperature as defined by
$T_F=E_F/k_B$, and $k_B$ is Boltzmann's constant. The Fermi energy,
$E_F=h\cdot (9.4\pm 0.5$ kHz), is determined from a measurement of
the peak density of the trapped gas. For the photoemission
spectroscopy, we apply an rf pulse that couples atoms in one of the
two spin states to an unoccupied third spin state. There are two
essential requirements for determining the excitation spectrum. The
first is that the interaction energy is sufficiently small that
$\epsilon_k =\hbar^2 k^2/2m$ holds and the data are not subject to
complicated final-state effects \cite{Chin2005, Yu2006, Punk2007,
Perali2008, Basu2007, Veillette2008, Levin2008}. The second
requirement is that collisions do not scramble the energy and
momentum information carried by the out-coupled atoms. In previous
rf spectroscopy measurements both of these requirements were not
satisfied \cite{Chin2004a, Gupta2003, Shin2007, Schunck2007,
Schunck2008}. In our $^{40}$K gas, however, the interaction energy
for the out-coupled atoms is approximately 640 Hz, which is much
smaller than $E_F$. Furthermore, the mean-free path for the
out-coupled atoms is much larger than the size of the gas:
$\frac{1}{\sigma n} \approx 6 \hspace{1mm}R_F$, where $\sigma$ is
collision cross section, $n$ is the average density, and $R_F$ is
the Fermi radius of the non-interacting gas.

\begin{figure}
\includegraphics[width=\linewidth]{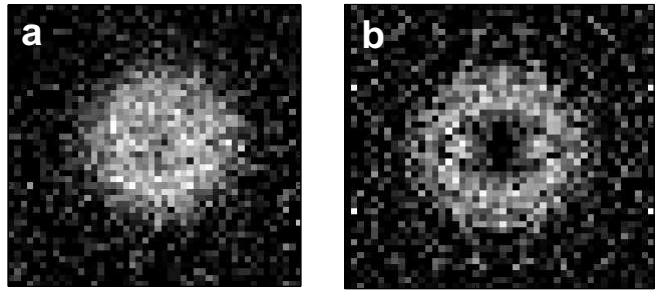}
\caption{\label{absimage} \textbf{Extracting the 3D momentum
distribution}. \textbf{a} A time-of-flight absorption image ($ \sim
145\mu$m$\times 145 \mu $m) of atoms that have been transferred into
a third spin state is taken after applying an rf pulse to a Fermi
gas on the BEC side of the Feshbach resonance. \textbf{b} After
performing quadrant averaging we use an inverse Abel transform to
reconstruct the 3D momentum distribution. For this particular
example, a 2D slice at the center reveals a shell-like structure for
the momentum distribution of the out-coupled atoms.}
\end{figure}

To resolve the kinetic energy, $\epsilon_k$, of the rf out-coupled
atoms we apply an rf pulse that is short compared to the trap
period. We then immediately turn off the trap, let the gas
ballistically expand, and measure the velocity distribution using
state-selective time-of-flight absorption imaging, see Fig.
\ref{absimage}. Assuming a symmetric momentum distribution, we
extract the 3D momentum distribution of the out-coupled atoms from
the 2D image by performing an inverse Abel transform.

\begin{figure}
\includegraphics[width=\linewidth]{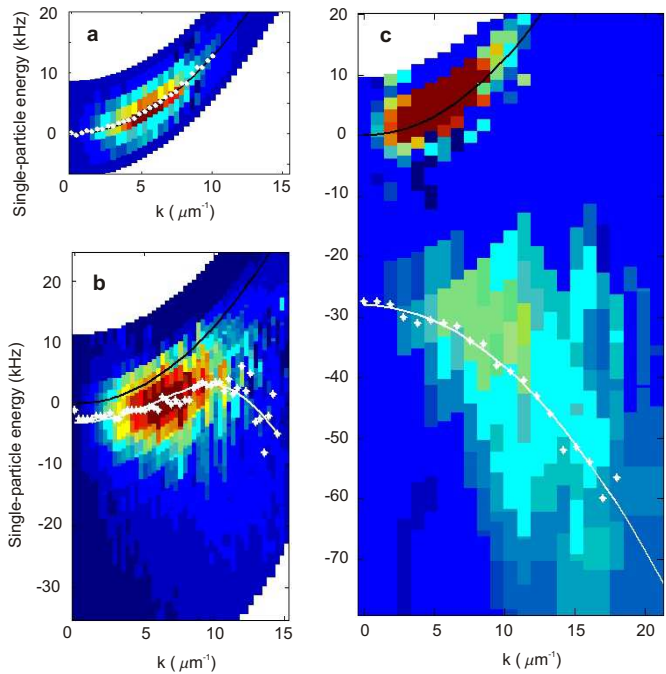}
\caption{\label{threedispers} \textbf{Single-particle excitation
spectra obtained using photoemission spectroscopy for ultracold
atoms}. Plotted are intensity maps (independently scaled for each
plot) of the number of atoms out coupled to a weakly-interacting
spin state as a function of the single-particle energy $E_s$ and
wave vector $k$. The black lines are the expected dispersion curve
for an ideal Fermi gas. The white points (*) mark the center of each
fixed energy distribution curve. \textbf{a} Data for a very
weakly-interacting Fermi gas. The Fermi wave vector $k_F^0$ is $8.6
\pm 0.3$ $\mu m^{-1}$. \textbf{b} Data for a strongly interacting
Fermi gas $1/k_F^0a=0$ and $T \approx T_c$.  The white line is a fit
of the centers to a BCS-like dispersion. \textbf{c} Data for a gas
on the BEC side of the resonance where $1/k_F^0 a \approx 1$ and the
measured two-body binding energy is $h\cdot (25\pm2$ kHz). We
attribute the upper feature to unpaired atoms and the lower feature
to molecules. The white line is a fit to the centers using a
quadratic dispersion.}
\end{figure}

We first consider the case of an ideal Fermi gas. To create a very
weakly interacting gas we adiabatically ramp the magnetic field to
the zero crossing of the Feshbach resonance. In Fig.
\ref{threedispers}a, we plot the intensity, which is proportional to
the number of atoms transferred into the third spin state, as a
function of the original single-particle energy $E_s$ and wave
vector $k$. The data are obtained by varying the rf frequency and
counting the out-coupled atoms as a function of their momenta. We
define zero energy to be the energy of a non-interacting atom at
rest in the initial spin state. The intensity map for a
non-interacting Fermi gas is expected to show delta function peaks
at $E_s=\epsilon_k$.  The white asterisks mark the centers of the
intensity at each value of $k$ as determined from Gaussian fits;
these show good agreement with the expected dispersion (black line).
The rms width in $E_s$ of the measured spectrum in Fig.
\ref{threedispers}a is 2.1 kHz and is due to an energy resolution
that comes from the rf pulse duration.

\begin{figure}
\includegraphics[width=\linewidth]{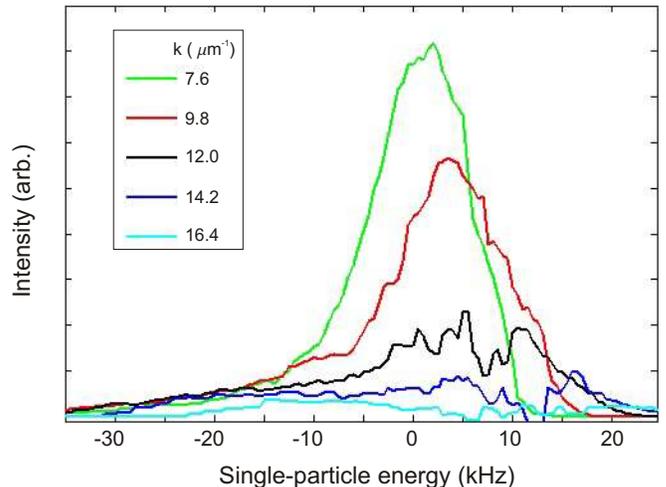}
\caption{\label{onresEDCs} \textbf{Energy distribution curves for a
strongly interacting Fermi gas}. We plot the intensity for selected
values of $k$. Each curve is the average for seven neighboring
values of $k$ in Fig. \ref{threedispers}b. The data have been
smoothed with a 1.5 kHz wide filter.}
\end{figure}

To create a strongly interacting Fermi gas we adiabatically ramp the
magnetic field to the peak of the Feshbach resonance where the
s-wave scattering length $a$ diverges and the dimensionless
interaction parameter $1/k_F^0a=0$. Here $k_F^0$ is the Fermi wave
vector that corresponds to the peak density of the original weakly
interacting gas. Previous measurements have shown that after the
ramp to $1/k_F^0a=0$, our Fermi gas will be at $(0.9\pm0.1)\cdot
T_c$ for the superfluid state \cite{Regal2004a}. With photoemission
spectroscopy on the strongly interacting gas we extract the
intensity map shown in Fig. \ref{threedispers}b. The interactions
lower the overall energy and flatten the dispersion curve. In
addition, the energy width is broadened well beyond our energy
resolution.

There is now a wide consensus that interpretation of previous rf
spectroscopy measurements \cite{Chin2004a, Shin2007, Schunck2007} in
terms of a pairing gap is a difficult problem that is still unsolved
theoretically \cite{Giorgini2008}. The photoemission spectroscopy
technique presented here directly measures the occupied
single-particle density of states and is therefore well-suited for
measuring pairing gaps.  In BCS theory the gap vanishes at $T_c$;
however, in the BCS-BEC crossover a pseudogap due to preformed pairs
is predicted to exist above $T_c$ \cite{Perali2002,Levin2005}.
Perali \emph{et al.} calculated the spectral function for a
homogeneous Fermi gas near $T_c$ and found that the peaks of the
spectral function fit almost exactly to a ``BCS-like" dispersion
curve where the BCS gap was replaced by the pseudogap
\cite{Perali2002}. As a first step to analyzing our data, we fit the
centers of the intensity at each value of $k$ to this BCS-like
dispersion curve \cite{Perali2002}, $E_s = \mu'
-\sqrt{(\epsilon_k-\mu')^2 +\Delta^2}$. Here, the fitting parameters
are the renormalized chemical potential $\mu'$ and the pseudogap
$\Delta$. The best fit, shown as the white curve in Fig.
\ref{threedispers}b, gives $\mu'=h \cdot (12.6\pm0.7$ kHz) and
$\Delta=h\cdot (9.5\pm0.6$ kHz).  In Fig. 14 of Ref.
\cite{Perali2002}, Perali \emph{et al.} also plot an example of
predicted spectral functions for a few values of wave vector $k$. To
facilitate comparison with theory, in Fig. \ref{onresEDCs} we show
measured energy distribution curves (EDCs) for select values of $k$.
It should be noted that in all trapped gas experiments, the density
is inhomogeneous and the pairing gap will depend on the local Fermi
energy. Therefore, our data should eventually be compared with a
theory that includes the effect of the trapping potential through,
for example, a local density approximation. Finally, we note that we
have performed photoemission spectroscopy for a gas cooled below
$T_c$ (initial $T/T_F = 0.10$) and found that the data is
qualitatively very similar to that in Fig. \ref{threedispers}b.

Far on the BEC side of the resonance, for $1/k_F^0 a \gg 1$, the
pairing gap eventually becomes a two-body binding rather than a
many-body effect that depends on the local Fermi energy. We measure
the excitation spectrum for the Fermi gas at $1/k_F^0 a \approx 1$
where the molecule binding energy measured for a low density gas is
$h\cdot(25\pm2$ kHz). We observe two prominent features, see Fig.
\ref{threedispers}c. The first feature is narrow in energy, starts
at zero energy, and follows the quadratic dispersion expected for
free atoms (black line). We attribute this feature to unpaired
atoms, which may be out of chemical equilibrium with the pairs. The
second feature is very broad in energy, is shifted to lower energy,
and trends towards lower energy for increasing $k$. This feature we
attribute to atoms in the paired state. An excitation gap separating
the two features is evident in the data.  We fit the centers of the
molecule feature to a quadratic dispersion (white line) with the
free parameters being the energy offset and an effective mass $m^*$.
In the BEC limit, where one has tightly bound molecules, we would
expect the energy offset to be the molecule binding energy, which
equals $2\Delta$, and the effective mass to be $-m$. This negative
effective mass reflects the fact that out coupling an atom at high
$k$ leaves behind an excitation in the form of an unpaired atom. The
best fit to the data gives an energy offset of $28$ kHz and
$m^*=-1.25\hspace{1mm}m$.

\begin{figure}
\includegraphics[width=\linewidth]{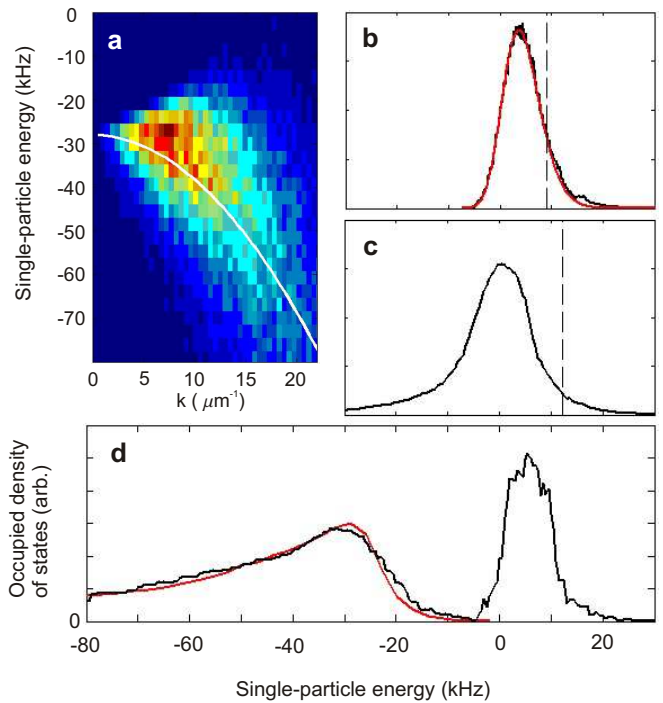}
\caption{\label{threeDOS} \textbf{The occupied single-particle
density of states}. \textbf{a} A calculated intensity map for a
thermal distribution of weakly bound molecules is shown. The white
line is the fit to the data shown in Fig. \ref{threedispers}c.
\textbf{b} The density of states for a weakly-interacting Fermi gas
(black line) agrees well with a fit (red curve) for a Fermi gas in a
harmonic trap. The fit, whose only free parameter is the amplitude,
includes our measurement resolution. The dashed black vertical line
shows $E_F$. \textbf{c} The density of states taken at the peak of
the Feshbach resonance is shifted to much lower energy. \textbf{d}
The density of states on the BEC side of the resonance has two
features; a narrow peak due to unpaired atoms and a broader feature
due to molecules. The red curve is the expected density of states
from the simulation shown in \textbf{a}.}
\end{figure}

The large energy width seen in Fig. \ref{threedispers}c is likely
due to center-of-mass motion of the pairs.  For comparison with the
data, we have performed a simple Monte Carlo simulation assuming a
thermal distribution for the center-of-mass motion and using the
predicted distribution of relative kinetic energy for rf
dissociation of weakly bound molecules \cite{Chin2005}.  We assume
the pairs are in thermal equilibrium with the unpaired atoms and use
the measured temperature of the rf out-coupled atoms corresponding
to the upper feature in Fig. \ref{threedispers}c.  Assuming a
molecule binding energy of $h\cdot(25$ kHz), the calculation gives
the intensity map shown in Fig. \ref{threeDOS}a.

The occupied density of states is obtained by summing the data in
Fig. \ref{threedispers} over all $k$, see Fig. \ref{threeDOS}b-d.
For the nearly ideal Fermi gas data, Fig. \ref{threeDOS}b, we find
good agreement with the expected density of states for a Fermi gas
at $T=0.18 \hspace{1mm}T_F$ in a harmonic trap (red curve). For the
strongly interacting gas, Fig. \ref{threeDOS}c, the occupied density
of states becomes wider in energy and the peak shifts towards lower
energies by an amount comparable to $E_F$. From previous
measurements of the in-trap size of the cloud \cite{Stewart2006} we
estimate the Fermi energy of the strongly interacting gas to be $h
\cdot (12.4\pm 0.7$ kHz) (dashed line). For the BEC side of the
resonance, Fig. 5d, a pairing gap between bound pairs and free atoms
is readily apparent. The red curve is the expected density of states
determined from the simulation of a thermal distribution of weakly
bound molecules (Fig. \ref{threeDOS}a). The only free parameter in
the simulation is an overall scaling factor.

In this work, we have used photoemission spectroscopy, accomplished
by momentum resolving the out-coupled atoms in rf spectroscopy, to
probe the occupied single-particle density of states and energy
dispersion through the BCS-BEC crossover. In the future, it may be
possible to use spatially resolved photoemission spectroscopy to
probe the local pairing gap. Another extension of this work will be
to study the BCS-BEC crossover as a function of temperature and/or
unbalanced spin population. Photoemission spectroscopy for ultracold
atoms is a powerful and conceptually simple probe of strongly
correlated atom gases that could be applied to many other atom gas
systems. In the studies presented here, the atoms are interacting
via isotropic s-wave interactions and therefore considering
different directions of the out-coupled atoms' momenta was not
necessary. However, like angle-resolved photoemission spectroscopy
(ARPES) for solids, this technique could also be applied to
non-isotropic systems such as atoms in an optical lattice, low
dimensional systems, or higher partial wave pairing of atoms
\cite{Gaebler2007}.

\subsection{Methods}
We evaporatively cool an equal mixture of $^{40}$K atoms in the
$|f,m_f \rangle =|9/2,-7/2 \rangle$ and $|f,m_f \rangle =|9/2,-9/2
\rangle$ states, where $f$ and $m_f$ give the hyperfine level in the
ground-state manifold. We cool the gas to in an optical dipole trap
as described previously \cite{Stewart2006}. The frequencies for the
cylindrically symmetric trap are $f_r= 233$ Hz and $f_z = 19$ Hz.
The time-of-flight imaging uses a beam that propagates along the
$\hat{z}$ direction. At the end of the evaporation, we adiabatically
increase the interaction strength by lowering the magnetic field, at
a rate of 0.1 G/ms, to a value near the Feshbach resonance located
at 202.10 $\pm$ 0.07 G \cite{Regal2004a}. The magnetic field values
for the data in Fig. \ref{threedispers}a-c are 208.43, 202.10, and
201.51 G, respectively.

For photoemission spectroscopy, we apply an rf pulse with a Gaussian
amplitude envelope with a $1/e^2$ width of $240\hspace{1mm}\mu$s, to
transfer atoms from the $|9/2,-7/2 \rangle$ state to the $|9/2,-5/2
\rangle$ state. The rf frequency is approximately 47 MHz. Atoms in
the $|9/2,-5/2 \rangle$ state have a two-body s-wave scattering
length that is 130 Bohr radii with the $|9/2,-7/2 \rangle$ state and
250 Bohr radii with the $|9/2,-9/2 \rangle$ state. Immediately after
the rf pulse we turn off the optical trap and let the atoms expand
for 3 to 6.5 ms before taking a resonant absorption image of the
$|9/2,-5/2 \rangle$ atoms. Typically four images are averaged for
each rf frequency.  For the weakly-interacting Fermi gas, the rf
power was chosen to achieve maximum transfer. For data on resonance
and on the BEC-side of resonance, no more than 30\% of the atoms
were transferred into the third spin state. At the higher rf
frequencies, we increased the signal by increasing the rf power and
then scaled the data to give the appropriate intensity.

In Eqn. \ref{eq:ConservE}, we take $\phi$ to be the rf resonance
energy measured for a weakly interacting gas. There is an
uncertainty of $\pm$ 1 kHz in $\phi$ and therefore also in the zero
of $E_s$. The measured wave vector $k$ has an uncertainty of
approximately 5\% from uncertainty in the magnification of the
imaging system.

\begin{acknowledgments}
We acknowledge funding from the NSF.  We thank Eric Cornell, Dan
Dessau, and the JILA BEC group for discussions.
\end{acknowledgments}

% You should use BibTeX and apsrev.bst for references
% Choosing a journal automatically selects the correct APS
% BibTeX style file (bst file), so only uncomment the line
% below if necessary.

%\bibliographystyle{prsty}
%\bibliography{Mom_resolved}

\end{document}